\documentclass[article,12pt,onecolumn,showpacs,preprintnumbers,amsmath,assume,plain]{revtex4}
\usepackage{graphicx}
\usepackage{dcolumn}
\usepackage{bm}
\usepackage{hyperref}

\def\bk{ {\bf k} }

\def\bd{ {\bf d} }

\def\sr{ Sr$_2$RuO$_4$ }

\begin{document}

\title{Anisotropic shear stress $\sigma_{xy}$ effects in the basal plane of \sr}
\author{P. Contreras $^{1}$}
\affiliation{$^{1}$ Departamento de F\'{\i}sica and Centro de
F\'{\i}sica Fundamental, Universidad de Los Andes, M\'erida
5101, Venezuela}
\author{Juan Moreno $^{2}$}
\affiliation{ $^{2}$ Facultad de Ingenier\'{\i}a, Universidad
de la Empresa, Montevideo 11300, Uruguay}
\date{\today}

\begin{abstract}
In this short note following a previous set of works
\protect\cite{walk,wak1,con1,con2} we repeat the calculations
for the jumps for the thermal expansion $\alpha_{\sigma_{xy}}$,
the specific heat C$_{\sigma_{xy}}$, and the elastic compliance
S$_{xyxy}^{\sigma_{xy}}$ in the basal plane of \sr. We use here
the $4^{th}$ rank tensor notation because of the Voigt
notation; where the stress and strain are treated differently.
Henceforth, we clarify some issues regarding a Ginzburg-Landau
analysis suitable to explain the sound speed experiments
\protect\cite{lup2}, and partially the strain experiments
\protect\cite{hick,tep1} in strontium ruthenate. We continue to
propose that the discontinuity in the elastic constant
C$_{xyxy}$ of the tetragonal crystal \sr gives an unambiguous
experimental evidence that the \sr superconducting order
parameter $\Psi$ has two components with a broken time-reversal
symmetry state, and that the $\gamma$ band couples the
anisotropic electron-phonon interaction to the $[xy]$ in-plane
shear stress in \sr \protect\cite{wak1,con1}. Some important
words about the roll of the spin equal to one for transversal
phonos are added following Levine [34].


\end{abstract}


\maketitle

\section{Introduction}\label{sec:intro}

In \sr the electrons in the Cooper pairs are bound in spin
triplets, where the spins are lying on the basal plane and the
pair orbital momentum is directed along the z-direction.
Henceforth, the order parameter $\Psi$ is represented by a
vector $\bd (\bk)$ (of the type $k_x \pm i\:k_y$)
\protect\cite{mae94,mae1,mac1,ric2}.

Based on the results of the Knight-shift experiment performed
through T$_c$ \protect\cite{ish1,duf1}, it has been proposed
that \sr is a triplet superconductor, also it has been reported
\protect\cite{luk1} that $\Psi$ breaks time-reversal symmetry,
which constitutes another key feature of unconventionality. The
\sr elastic constants C$_{xyxy}$ have been carefully measured
as the temperature T is lowered through T$_c$, showing the
existence of small step in the transverse sound mode $T[100]$
\protect\cite{lup2}. This experimental result theoretically
implies that $\Psi$ has two different components with a
time-reversal symmetry broken state \protect\cite{con2}.
Similar conclusions from a muon-spin relaxation ($\mu$SR)
experiment were reported \protect\cite{luk1}. Recently,
experiments on the effects of uniaxial strain $\epsilon_{xy}$,
were performed \protect\cite{hick}, reporting that for \sr the
symmetry-breaking field can be controlled experimentally.

Additionally, a most recently experiment \protect\cite{tep1}
found that the transition temperature T$_c$ in the
superconductor \sr rises dramatically under the application of
a planar anisotropic strain, followed by a sudden drop beyond a
larger strain. Furthermore, recent theoretical work suggests
that those recent experiments tuned the Fermi surface topology
efficiently by applying planar anisotropic strain emphasizing
again, the point of view that in-plane effects (even by means
of a more complicated renormalization group theory framework)
also shows clear evidence of a symmetry broken stated in \sr.
Furthermore, they reported a rapid initial increase in the
superconducting transition temperature T$_c$, that the authors
associated with the evolution of the Fermi surface toward a
Lifshitz Fermi surface reconstruction under an increasing
strain \protect\cite{liu}.

Here, we aim to clarify some particular concepts and methods
following an elastic phenomenological (GL) approach
\protect\cite{lan1,walk,lan2,tes1,aul1,mus,jab1}. First, the
differences between using stress or strain (which is the
response of a system to applied stress) \protect\cite{grechka}
to explain a time-reversal symmetry broken state. Second, there
is a claim \protect\cite{liu} that only the $\gamma$ band
responses to the strain sensitively, and we emphasize that this
physical phenomenon is caused by the $\gamma$ band coupling the
anisotropic electron-phonon interaction to the $[xy]$ plane
\protect\cite{wak1,con1}. Third, we do not expand our analysis
to a Lifshitz reconstruction of the Fermi surface mainly
because we do not have experimental evidences that show a
topological Lifshitz  transition in \sr even in its normal
state, neither we have observed a generalized topological
transition in \sr. In our opinion a 2D Fermi contour evolution
under an applied external strain as the one for the $\gamma$
band in \sr needs further interpretations in terms of a
topological electronic Lifshitz phase transition
\protect\cite{lif,kag1,kag2,kag3}. We remember that
$\sigma_{ik} = - p \; \delta_{ik}$ shows how pressure and
stress are in general related, the stress $\sigma_{ik}$ becomes
the delta function $\delta_{ik}$ if a volumetric pressure is
applied to a sample \protect\cite{lan4}.

These experimental results and theoretical interpretations need
to clarify moderately in order to unify several theoretical
criteria which try to explain the changes occurring in the
C$_{xyxy}$ elastic constant at T$_c$
\protect\cite{lup2,hick,tep1}.

Thus, the aim of this short note is to clarify again that an
elastic Ginzburg-Landau phenomenological approach partially
demonstrates that \sr is an unconventional superconductor with
a two-component order parameter $\Psi$
\protect\cite{con1,con2}. We based our interpretation on a \sr
$[T100]$ transversal response-impulse mode experimentally
measured as the temperature T is lowered through T$_c$ showing
only an small step change \protect\cite{lup2} (for that
particular result please see bottom panel on Fig 5.7 and Fig
5.8 pp. 138-139 \protect\cite{lup2}). The result clearly shows
a discontinuity in the $T[100]$ mode.

\begin{figure}[htp]
\begin{center}
\includegraphics[width = 3.0 in, height= 2.0 in]{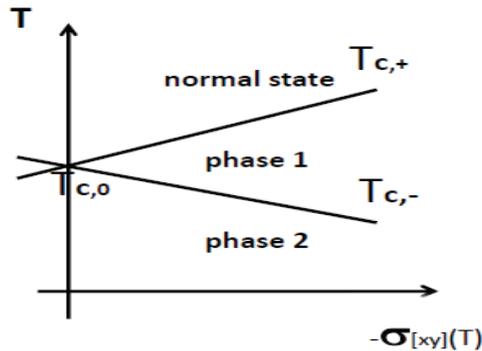}
\end{center}
\caption{\label{GL3} Phase diagram sketch showing the upper $T_{c+}$, lower $T_{c-}$,
and zero $T_{c0}$ superconducting transition temperatures as
a function of the shear stress $-\sigma_{xy}(T)$. Notice that at $T_{c0}$ the derivative
$\frac{d \; T_{c0}}{d \; \sigma_{xy}}$
does not exist.}
\end{figure}

Here, let us mention that a different theory of \sr elastic
properties was presented \protect\cite{sig2}. However, the
approach followed does not take into account the splitting of
T$_c$ due to shear $\sigma_{xy}$, and directly calculates the
jumps at zero stress, where the derivative of T with respect to
$\sigma_{xy}$ does not exist \protect\cite{walk} (see
fig.\ref{GL3}).

\section{Shear stress $\sigma_{xy}$ analysis}

\begin{figure}[htp]
\begin{center}
\includegraphics[width = 3.0 in, height= 2.0 in]{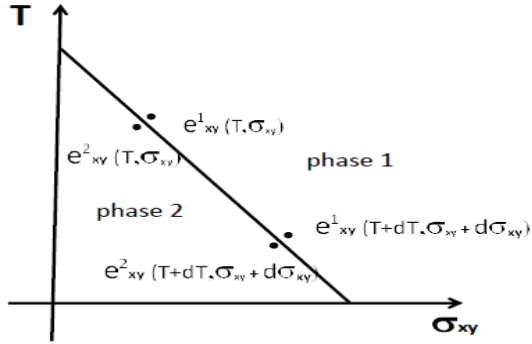}
\end{center}
\caption{\label{GL5} Sketch showing the line of transition temperatures T$_c$ along of a second order phase transition.
The state of the strain e$_{xy}$ (shown in the figure), the entropy $S$, and the volume $V$
are continuous functions along the line of the second order phase transition \protect\cite{walk,con1}. }
\end{figure}

In this section, we make use of the $4^{th}$ rank tensor
notation because the Voigt notation has a disadvantage; the
stress and strain are treated differently. Voigt mapping only
preserves the elastic stiffnesses. We also call the uniaxial
shear stress as shear stress only because the effect observed
is in the basal plane $[xy]$. As was stated previously
\protect\cite{con1,con2} when shear stress $\sigma_{xy}$ is
applied to the basal plane of \sr, the crystal tetragonal
symmetry is broken, and a second-order transition to a
superconducting state occurs. Accordingly, for this case the
analysis of the sound speed behavior at T$_c$ also requires a
systematic study of the two successive second order phase
transitions (see fig.\ref{GL5}). Hence, the C$_{xyxy}$
discontinuity \protect\cite{lup2} at T$_c$, can be explained in
this context. Due to the absence of discontinuity in S$_{xyxy}$
for any of the one-dimensional $\Gamma$ representations, the
superconductivity in \sr must be described by the two
dimensional irreducible representation E$_{2u}$ of the
tetragonal point group D$_{4h}$ \cite{con2}.

If there is a double transition, the derivative of T$_c$ with
respect to $\sigma_{xy}$ i.e. $d T_c/d \sigma_{xy}$ is
different for each of the two transition lines (see the
T$_c$-$\sigma_{xy}$ phase diagram in fig.\ref{GL5}). At each of
these transitions, the specific heat C$_{\sigma_{xy}}$, the
thermal expansion $\alpha_{\sigma_{xy}}$, and the elastic
compliance S$_{xyxy}^{\sigma_{xy}}$ show discontinuities
\protect\cite{con2}, the sum of them gives the correct
expressions for the discontinuities at zero shear stress, where
the Ehrenfest relations do not hold.

In the case of an applied shear stress $\sigma_{xy}$, the
change for the in-plane Gibss free energy is given by
\begin{eqnarray}
   \Delta G_{\sigma_{xy}} & = &\alpha ( |\psi_x|^2 + |\psi_y|^2 ) + \sigma_{xy} d_{xyxy} (\psi_x \psi^{\ast}_y + \psi^{\ast}_x \psi_y) +
   \nonumber\\
   & & \frac{b_1}{4}(|\psi_x|^2 +|\psi_y|^2)^2 + b_2|\psi_x|^2 |\psi_y|^2 +
   \nonumber \\
   & & \frac{b_3}{2}(\psi_x^2 \psi_y^{*2} + \psi_y^2 \psi_x^{*2}),
   \label{sts_1}
\end{eqnarray}
where $d_{xyxy}$ term couples the stress $\sigma_{xy}$ to the
order parameter, the thermal expansion coefficient $\alpha$ =
$\alpha^\prime (T - T_{c0})$, and the minimization of $\Delta
G_{\sigma_{xy}}$ is performed by substituting the general
expression for $\Psi$ as was previously calculated
\protect\cite{con1,sig2}. Therefore $\Delta G_{\sigma_{xy}}$
becomes
\begin{eqnarray}
    \Delta G_{\sigma_{xy}} & = & \alpha (\eta_x^2 + \eta_y^2) + 2 \eta_x \eta_y \: \sigma_{xy} \sin \varphi
    \: d_{xyxy} + \frac{b_1}{4}(\eta_x^2 +\eta_y^2)^2 + \nonumber \\
    & & (b_2 - b_3)\eta_x^2 \eta_y^2 + 2 b_3 \eta_x^2 \eta_y^2 \sin^2 \varphi.
    \label{st_ss3}
\end{eqnarray}

In the presence of $\sigma_{xy}$, the second order term
determines the phase below T$_{c+}$, which is characterized by
$\psi_x$ and by $\psi_y$ $=$ 0. As the temperature is lowered
below T$_{c-}$, depending of the value of b$_3$ a second
component $\psi_y$ may appear. If at T$_{c-}$ a second
component occurs, the fourth order terms in eqn.~(\ref{st_ss3})
will be the dominant one. Thus for very low T's, or for
$\sigma_{xy}$ $\rightarrow$ 0, a time-reversal
symmetry-breaking superconducting state may emerge. The
analysis of eqn.~(\ref{st_ss3}) depends on the relation between
the coefficients b$_2$ and b$_3$. It also depends on the values
of the quantities $\eta_x$ and $\eta_y$, and of the phase
$\varphi$. If b$_3$ $<$ 0, and $\eta_x$ and $\eta_y$ are both
nonzero, the state with minimum energy has a phase $\varphi$ =
$\pi/2$. The transition temperature is obtained from
eqn.~(\ref{st_ss3}), by performing the canonical
transformations: $\eta_x$ = $\frac{1}{\sqrt{2}}(\eta_{\mu} +
\eta_{\xi})$ and $\eta_y$ = $\frac{1}{\sqrt{2}}(\eta_{\mu} -
\eta_{\xi})$. After their substitution, eqn.~(\ref{st_ss3})
becomes
\begin{equation}
    \Delta G_{\sigma_{xy}} = \alpha_{+} \eta_{\xi}^2 + \alpha_{-} \eta_{\mu}^2  +
    \frac{1}{4}(\eta_{\xi}^2 + \eta_{\mu}^2)^2 + (b_2 + b_3)(\eta_{\xi}^2 - \eta_{\mu}^2)^2.
    \label{st_ss5}
\end{equation}

As it was before done, $\eta_{\xi} = \eta \hspace{0.1cm} \sin
\chi$ and $\eta_{\mu} = \eta \hspace{0.1cm} \cos \chi$,
eqn.~(\ref{st_ss5}) takes the form
\begin{equation}
    \Delta G_{\sigma_{xy}} = \alpha_{+} \eta^2 \sin^2\chi + \alpha_{-} \eta^2 \cos^2\chi  +
    \frac{\eta^4}{4}\Big[ b_1 + (b_2 + b_3) \cos^2 2\chi \Big].
    \label{st_ss6}
\end{equation}

$\Delta G_{\sigma_{xy}}$ is minimized if $\cos 2\chi $ = 1,
this is, if $\chi$ $=$ 0. Also, in order for the phase
transition to be of second order, $b^\prime$, defined as
$b^\prime \equiv b_1 + b_2 + b_3$, must be larger than zero.
Therefore, if $\sigma_{xy}$ is non zero, the state with the
lowest free energy corresponds to $b_3$ $<$ 0, phase $\varphi$
equal to $\pi/2$, and $\Psi$ of the form:
\begin{equation}
(\psi_x,\psi_y) \approx \eta \: (e^{\frac{i \varphi}{2}}, \: e^{-\frac{i \varphi}{2}}).
\end{equation}

In phase 1 of fig.~(\ref{GL5}), $\varphi$ = 0, and as T is
lowered below T$_{c-}$, phase 2, $\varphi$ grows from 0 to
approximately $\pi/2$. The two transition temperatures T$_{c+}$
and T$_{c-}$ are obtained to be:
\begin{eqnarray}
T_{c+}(\sigma_{xy}) & = & T_{c0} - \frac{\sigma_{xy}}{\alpha^\prime} d_{xyxy}, \\
T_{c-}(\sigma_{xy}) & = & T_{c0} + \frac{b}{2 \: b_3 \:
\alpha^\prime} \sigma_{xy} \; d_{xyxy}. \label{st_s7}
\end{eqnarray}

The derivative of T$_{c+}$ with respect to $\sigma_{xy}$, and
the discontinuity in $C^+_{\sigma_{xy}}$ at T$_{c+}$  are
respectively:

\begin{eqnarray}
\frac{d \; T_{c+}}{d \; \sigma_{xy}} & = & - \frac{d_{xyxy}}{\alpha^\prime}, \\
\Delta C^+_{\sigma_{xy}} & = & -\frac{2 \: T_{c+} \:
\alpha^{\prime 2}}{b^\prime}. \label{st_s81}
\end{eqnarray}

After applying the Ehrenfest relations \protect\cite{lan1} the
results for $\Delta \alpha_{\sigma_{xy}}$ and $\Delta
\text{S}_{xyxy}$ at T$_{c+}$ are:
\begin{eqnarray}
\Delta \alpha_{\sigma_{xy}}^+ & = & -\frac{2 \: \alpha^\prime \: d_{xyxy}}{b^\prime}, \\
\Delta \text{S}_{xyxy}^+ & = & -\frac{2 \:
d_{xyxy}^2}{b^\prime}. \label{st_s82}
\end{eqnarray}

For T$_{c-}$, the derivative of this transition temperature
with respect to $\sigma_{xy}$, and the discontinuities in the
specific heat, thermal expansion and elastic stiffness
respectively are:
\begin{eqnarray}
\frac{d \; T_{c-}}{d \; \sigma_{xy}} & = &  \frac{b \: d_{xyxy}}{2 \: b_3 \: \alpha^\prime}, \\
\Delta C_{\sigma_{xy}}^- & = & -\frac{4 \: T_{c-} \:
\alpha^{\prime 2} \: b_3}{b \: b^\prime}, \label{st_s91}
\end{eqnarray}

\begin{eqnarray}
\Delta \alpha_{\sigma_{xy}}^- & = & \frac{2 \: \alpha^\prime \: d_{xyxy}}{b^\prime}, \\
\Delta \text{S}_{xyxy}^- & = & -\frac{d_{xyxy}^2 \: b}{b^\prime
b_3}. \label{st_s92}
\end{eqnarray}

\begin{figure}[ht]
\begin{center}
\includegraphics[width = 3.0 in, height= 2.0 in]{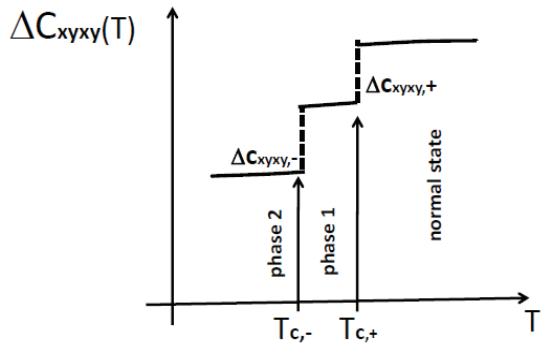}
\end{center}
\caption{\label{EC} Schematic dependence of the elastic constant on
the temperature for Sr$_2$RuO$_4$. Notice the two jumps in the
in plane elastic constant C$_{xyxy}$ near the transition temperatures T$_{c+}$ and
T$_{c-}$. The sketch shows two jumps of the same magnitude.
This happens for a two component order parameter $\Psi$.}
\end{figure}

\begin{figure}[ht]
\begin{center}
\includegraphics[width = 3.0 in, height= 2.0 in]{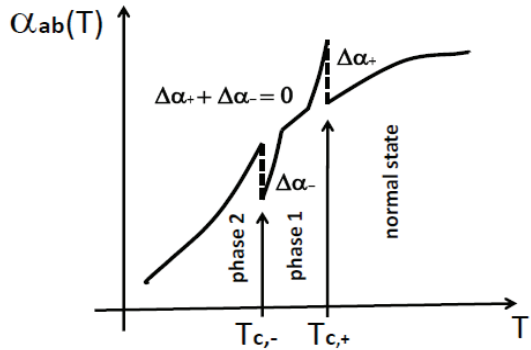}
\end{center}
\caption{\label{TE} Schematic dependence of the thermal expansion on
the temperature for Sr$_2$RuO$_4$. Notice the two jumps in the
in plane thermal expansion coefficient near the transition temperatures T$_{c+}$ and
T$_{c-}$. The sketch shows two jumps of the same magnitude but they have opposite signs, and their
sum cancels out at T$_{c0}$. This happens for a two component order parameter $\Psi$.}
\end{figure}

Since for the case of $\sigma_{xy}$, the derivative of T$_c$
with respect to $\sigma_{xy}$ is not defined at zero stress
point (see fig.\ref{GL3}), the Ehrenfest relations do not hold
at T$_{c0}$. Thus, the discontinuities occurring at T$_{c0}$,
in the absence of $\sigma_{xy}$, are calculated by adding the
expressions obtained for the discontinuities at T$_{c+}$ and
T$_{c-}$ in elastic stiffness fig.~(\ref{EC}), elastic
compliances and thermal expansion are given by:
\begin{eqnarray}
\Delta C_{\sigma_{xy}}^0  & = &  -\frac{2 \: T_{c0} \: \alpha^{\prime 2}}{b}, \\
\Delta \text{S}_{xyxy}^0 & = & -\frac{d_{xyxy}^2}{b_3}, \\
\Delta \alpha_{\sigma_{xy}}^0  & = &  0, \label{st_s10}
\end{eqnarray}
in this case, there is no discontinuity for the thermal
expansion of \sr at zero stress $\alpha_{\sigma_{xy}}^0$ which
is another physical feature we previously predicted
~\protect\cite{con2} (see fig.\ref{TE}). Some experimental
studies on the changes in the thermal expansion coefficient
$\alpha_i$ below $T_c$ in \textbf{HTS} reported
~\protect\cite{asa} that an additive lattice jump was found to
appear spontaneously at $T_c$ for a high T$_c$ compound with
one-component order parameter.

Since the phase diagram was determined as a function of
$\sigma_{xy}$, rather than as a function of the shear strain
$\epsilon_{xy}$, (see fig.~\ref{GL3}), in this work, as in
refs.~\protect\cite{con2,con1}, we make use of the 6 $\times$ 6
elastic compliance matrix $\text{S}$, and also of the full
range tensor notation. However, the sound speed measurements
~\protect\cite{lup2,lup1} are best interpreted in terms of the
elastic stiffness tensor $\text{C}$, with matrix elements
C$_{ijkl}$, which is the inverse of $\text{S}$
\protect\cite{nei1}. However, the strain is easier to measure
than the stress because fiber optics yields dynamic fidelity of
a fraction of a nanostrain \protect\cite{grechka}.

Therefore, it is important to be able to obtain the
discontinuities in the elastic stiffness matrix in terms of the
elastic compliance matrix for the shear stress case. Thus,
close to the transition line we follow \protect\cite{walk}:
$\text{C}(T_c + 0^+)$ = $\text{C}(T_c - 0^+) + \Delta \;
\text{C}$ and $\text{S}(T_c + 0^+)$ = $\text{S}(T_c - 0^+) +
\Delta \text{S}$, where $0^+$ is positive and infinitesimal. By
making use of the fact that $\text{C}(T_c + 0^+) \;
\text{S}(T_c + 0^+)$ = $\hat{1}$, where $\hat{1}$ is the unit
matrix, it is shown that, to first order, the discontinuities
satisfy, $\Delta \text{C}$ $\approx$ $- \text{C} \: \Delta
\text{S} \: \text{C}$. In this manner, it is found that, for
instance at T$_{c+}$, the expressions that define the jumps for
the discontinuities in elastic stiffness and compliances, due
to an external stress, have either a positive or a negative
value. In this way, $\Delta \text{S}_{xyxy}$ are negative;
while $\Delta \text{C}_{xyxy}$ are positive.

\section{Remarks} \label{cuatro}

The most noteworthy outcome of  this short note is that  the
observation of a discontinuity in the elastic constant
C$_{xyxy}$ \protect\cite{lup1,lup2} is an evidence  that the
order parameter $\Psi$ in \sr has two components as the
theoretical $GL$ analysis predicts. Also the theoretical
indicator that the sum of the jumps $\Delta
\alpha_{\sigma_{xy}}^{+} + \Delta \alpha_{\sigma_{xy}}^{-} = 0$
for the in-plane thermal expansion coefficient
\protect\cite{con2} (see fig.\ref{TE}). Hence, the use of \sr
as a material in detailed studies of superconductivity
symmetry-breaking effects has significant advantages because is
described by a two-component order parameter. Nevertheless,
determining from \sr experimental measurements the magnitude of
the parameters in the Ginzburg-Landau model is complicated
\protect\cite{con2}.

In the experimental work of the sound velocity measurements
\protect\cite{lup2,lup1} in \sr a discontinuity in the behavior
for C$_{xyxy}$ below T$_{c0}$, without a significant change in
the sound speed slope as T goes below 1 Kelvin was understood
as a signature of an unconventional transition to a
superconducting phase \protect\cite{lup2,con2,sig2}. Thus, this
set of previous results, and other recent results
\protect\cite{hick,tep1,liu,acha}, considers \sr as a strong
candidate for a detailed experimental investigation of the
effects of a symmetry-breaking field by means of strain or
stress experimental measurements. We also suggest that an
in-plane thermal expansion measurement at zero uniaxial shear
stress might further clarify any previous interpretation.

However, according to Levine \protect\cite{lev}, transversal
phonons may have spin equal to one for certain point group
symmetries. It could be that it occurs to the D$_{4h}$
tetragonal group of strontium ruthenate, this could be the key
to understand superconductivity and the time-reversal symmetry
broken state in \sr. The possible fact that phonons in the jump
observed in the sound transversal mode $T[100]$ and $T[110]$
\protect\cite{lup1,lup2} have spin equal to one. As a
consequence it could be that these symmetry breaking strain or
stress fields effects are due the transversal spin waves in
certain point group symmetries in metallic solids. Some words
about the roll of the magnetic interaction should be added.

\newpage

\section*{Acknowledgments}
P. Contreras express his gratitude to Professors C. Lupien and
L. Taillefer at Universite de Sherbrooke from for stimulating
discussions regarding their experimental results several years
ago at the University of Toronto. We also acknowledge Dr.
Vladimir Grechka from Marathon Oil for clarifying certain
aspects regarding the use of the shear strain for experimental
purposes.

\end{document}